# ENERGY DISTRIBUTION IN EARTHQUAKES: A NON-EXTENSIVE APPROACH

**Oscar Sotolongo-Costa** [1,2]; **A. Posadas**[2,3]
(1) Department of Theoretical Physics, Havana University, Habana 10400, Cuba
(2) Department of Applied Physics, Almeria University, Almeria, Spain
(3) Andalusian Institute of Geophysics and Seismic Disasters Prevention, Almeria University, Almeria, Spain.

## ABSTRACT

By using the maximum entropy principle with Tsallis entropy and under the assumption that the gouge plays an active role in the triggering of earthquakes, we obtain a functional dependence for energy distribution function for earthquakes which fits very well with observations in the region of small energies. This distribution function is related to the size distribution function of fragments in the gouge.

## INTRODUCTION

The Gutenberg-Richter law (1) has motivated a mass of research. This is due to the importance of the knowledge of the energy distribution of earthquakes and its physical and practical implications. Some famous models like those of Burridge-Knopoff (2) and Olami-Feder-Christensen (3) have focused in the mechanical phenomenology of earthquakes through simples images which capture essentials of the nature and genesis of a seism: the relative displacement of tectonic plates or the relative motion of the hanging wall and footwall in a fault and also the existence of a threshold for a catastrophic release of energy in the system.

The irregular geometry of the profiles of the tectonic plates was highlighted in (4) using a geometric viewpoint to deduce the power law dependence of earthquake energy distribution with good results. The importance of a geometric viewpoint for this phenomenon has been also highlighted in (5), where an idealized representation of the fragmented core of a fault (gouge) was presented.

Today it is widely accepted that most earthquakes are originated by relative motion of fault planes, whereas the images to model this energy release are diverse. The standard earthquake picture usually assigns the cause of an earthquake to some kind of rupture or some stick-slip mechanism in which the friction properties of the fault play the determinant role. A review of these viewpoints and some generated paradoxes can be found in (6).

Though the gouge has been recognized as an important factor in the dynamics of the earth's crust, we think that its role is more active in small faults and we intend to corroborate this viewpoint in this paper.

In (5) the gouge was idealized to be a medium formed by circular disk-shaped pieces filling the space between two planes acting like bearings. Here, we present a more realistic approximation, considering that the surfaces of the tectonic plates are irregular and the space between them contains fragments of very diverse shape.

We will present the geometric image of what we will call a "gouge earthquake" and in next sections we will justify the attempt to find the size distribution function of the fragments. Finally, we will compare our theoretical results with the earthquakes registered in the South of the Iberian Peninsula (Spain). The Andalusian Seismic Network collects data and the analysis of its catalogue is extended for more than 20000 earthquakes (m < 5.5). In this study we analyze about 12000 earthquakes ranging from 0 magnitude to 2.5.

## IMAGE FOR EARTHQUAKES IN THE GOUGE

The abundance of models that manifest the catastrophic mechanical nature of earthquakes is expected since the mechanism of earthquake triggering, although not well known, reveals itself as a catastrophic one.

The irregularity of the profiles of the tectonic plates has been pointed out as a main cause of earthquakes and in (4) the Gutenberg-Richter law for large earthquakes was obtained on the basis of assuming a Brownian shape of the profiles and the hypothesis that energy release is proportional to the overlap interval between profiles. This is a very appealing explanation, but in this as in many other models the material between the fault planes is not considered. When it is, as in (5), it is considered more to deaden than to trigger earthquakes. Nevertheless, the irregularities of the fault planes can be combined with the distribution of fragments between them to develop a mechanism of triggering earthquakes, which can be essentially the same as in (4) owing to the irregular shape of the profiles and fragments.

Let us start from the situation illustrated in the figure 1: two irregular profiles (no predetermined shape is assumed) are able to slip as shown in the figure. The motion can be hindered not only by the overlapping of two irregularities of the profiles, but also by the eventual relative position of several fragments as illustrated in the figure between the points "a" and "b".

Tensions in the resulting structure accumulated until a displacement of one of the asperities or even its breakage in the point of contact with the gouge fragment leads to a relative displacement of the fault planes of the order of the side of the hindering fragment "r". Then, the eventual release of tensions, whatever be the cause, leads to such a displacement to the subsequent liberation of energy.

We assume this energy "ε" to be proportional to "r", so that the size distribution of the fragments in the gouge can reflect the energy distribution of earthquakes generated by this mechanism. It seems natural to accept the energy release by this mechanism should be smaller than that related with the collision and breakup of the rough surface of the tectonic plate or a large fault, which is a tougher mechanism, i.e. "small earthquakes" should be well described by this model.

**THE MODEL**

As already pointed out, the size distribution function of fragments in the gouge should reflect the energy distribution function of the "gouge earthquake". We can assume that the constant interaction and local breakage of the fault planes produce the fragments. So, a fragmentation process produces them. The process of fault slip can be considered to occur in a homogeneous fashion in all the deep of the fault so that in any plane transverse to the deep of the fault the situation is the same. Then, to deduce the size distribution function of the fragments we consider a two-dimensional frame as the one illustrated in figure 1. Then, our problem is to find the distribution of fragments by area.

The maximum entropy principle has proved to be useful in the study of the fragmentation phenomena (7), though in that work an important feature of fragmentation i.e. the presence of scaling in the size distribution of fragments was not obtained. We will also apply the maximum entropy principle but now with recourse to the Tsallis entropy instead of the Boltzmann one. The Tsallis entropy for our problem has the form:

$$S_q \;=\; k \;\; \frac{1 - \int p^q(\sigma) d\sigma}{q - 1} \tag{1}$$

where $p(\sigma)$ is the probability of finding a fragment of relative surface $\sigma$ referred to characteristic surface of the system, and $q$ is a real number. $k$ is Boltzmann's constant. It is easy to see that entropy reduces to the Boltzmann's one when $q \to 1$. The sum by all states in the entropy is here expressed through the integration in all sizes of the fragments up to the largest surface $\Sigma$ of the collection of fragments.

The Tsallis formulation involves the introduction of at least two constraints. The first one is the normalization of $p(\sigma)$:

$$\int_0^\Sigma p(\sigma) d\sigma \;=\; 1 \tag{2}$$

and the other is the "*ad hoc*" condition about the q-mean value, which in our case can be expressed as:

$$\int_0^\Sigma \sigma \; p^q(\sigma) d\sigma \;=\; << \sigma >>_q \tag{3}$$

This condition reduces to the definition of the mean value when $q \to 1$. More information concerning the constraints that can be imposed in the formulation can be seen in (9).

This formulation of the statistical physics, known as "non extensive" formulation, since this entropy is not additive, has proved to be very useful to describe phenomena in which Boltzmann statistics fails to give a correct explanation, specially when the spatial correlations cease to be short ranged (10). Fracture is a paradigm of such long-range interaction phenomenon, and we gave a formulation in terms of Tsallis statistics very recently (11). Then the problem is to find the extremum of $\frac{s_q}{k}$ subject to the conditions given by formulas 2 and 3. To simplify we will assume $<<\sigma>>_q = 1$. As will be seen this has no effect in the final result. Application of the method of Lagrange multipliers gives:

$$p(\sigma) = \frac{\left[1 + (1-q)\beta\sigma\right]^{1/(1-q)}}{Z(q,\Sigma)} \qquad (4)$$

where $\beta$ is one of the Lagrange multipliers to be determined from the conditions 2 and 3. $Z(q,\Sigma)$ is some kind of partition function given by:

$$Z(q,\Sigma) = \int\limits_0^\Sigma \left[1 + (1-q)\beta\sigma\right]^{1/(1-q)} d\sigma \qquad (5)$$

If we now introduce the proportionality of the released relative energy $\varepsilon$ with the linear dimension r of the fragments, as $\sigma$ scales with $r^2$, the resulting expression for the energy distribution function of the earthquakes due to this mechanism is:

$$p(\varepsilon) = \frac{\varepsilon}{Z(q,\Sigma)\left[1 + (1-q)\beta\,\varepsilon^2\right]^{1/(q-1)}} \qquad (6)$$

Hence, we have obtained an analytic expression which can be fitted to the energy distribution of earthquakes through $Z(q,\Sigma)$, $\beta$ and q. This was obtained from a simple model starting from first principles. No *ad hoc* hypothesis was introduced but the proportionality of "$\varepsilon$" and "r", which seems justified.

## DATA AND APPLICATION OF THE MODEL

The Andalusian Institute of Geophysics and Seismic Disaster Prevention compiled the earthquake catalogue used in this study. The earthquakes were observed by the Andalusian Seismic Network of the Andalusian Institute of Geophysics and Seismic Disaster Prevention that consists of more than 20 observational stations [12]. The analyzed area is the rectangular region between 35º and 38º North Latitude and between 0º and 5º West Longitude, hereinafter called the South of the Iberian Peninsula region. Figure 2 shows the epicenters of the earthquakes of the South of the Iberian Peninsula region catalogue (more than 20000 events). Error of the hypocentres' location in the x, y and z directions are more or less ±1 km, ±1 km and ±2 km, respectively [13]. The seismicity during the period from 1985 to 2000 may be considered normal, i.e., without major seismic events. The Gutenberg-Richter relation is satisfied in this data set, at least for earthquakes with magnitude greater than 2.5. The data is assumed to be free of observational bias and abnormal seismicity. The formula 6 was used to fit the result of about 12000 earthquakes with magnitude less than 2.5.

In the figure 3, it is shown the cumulative number of earthquakes of magnitude "m" (earthquakes with a magnitude equal or larger than "m", being "m" the logarithm of the relative energy "$\varepsilon$") in the region of small energy vs. "m".

The cumulative number was calculated as the integral from "ε" to "∞"of the formula 6 as $N(\varepsilon) = \dfrac{a'\,(1+b'\varepsilon^2)^{1-c'}}{2b'(c'-1)}$ with a', b' and c' constants. On the other hand $m \propto \log(\varepsilon)$ where m is the magnitude, so easily we get:

$$\log(\,N(>m)\,) = a + b \cdot \log(\,1 + c \cdot 10^{2m}\,) \tag{7}$$

With a, b and c as parameters to fit the data. A good fitting can be seen with the integral of formula 6, showing that the functional dependence here obtained is correct. This is a very interesting fact since it is known that the Gutenberg-Richter's fails to describe small earthquakes [14] and large earthquakes [15]. The above-obtained expression gives a good fitting precisely in the region of small energies. This is an argument in favor of the proposed image from previous sections: small earthquakes seem thus to be closely related with the interactions in the gouge.

## CONCLUSIONS

A functional dependence was obtained for the distribution of earthquakes produced by interactions in the gouge, starting from a non-extensive formulation of the maximum entropy principle (the Tsallis formulation). The active role of the material between the tectonic plates was in evidence with this model that seems to relate some earthquakes with the material of the gouge. Tsallis formulation is, as can be seen, determinant to obtain the energy distribution of small earthquakes. No a priori assumption about the fault profile or shape of the fragments was needed.

## ACKNOWLEDGEMENTS

This work was partially supported by the CICYT projects AMB97-1113-C02-02 and AMB99-1015-C02-02, DGESIC project HF1999-0129, the Alma Mater contest, Havana University. One of us (O.S.) is grateful to the Department of mathematical Sciences of Brunel University for kind hospitality and the Royal Society, London, for financial support and also to the Department of Applied Physics of Almeria University.

**FIGURE CAPTIONS**

Figure 1. An illustration of the relative motion of two irregular faults in the presence of material filling the space between them. Observe that this material may play the role of bearing or also of particles that hinder the relative motion of the plates as seen in the figure between the points a and b.

Figure 2. The Andalusian Institute of Geophysics and Seismic Disaster Prevention compiled the earthquake catalogue used in this study. The earthquakes were observed by the Andalusian Seismic Network of the Andalusian Institute of Geophysics and Seismic Disaster Prevention that consists of more than 20 observational stations. More than 20000 earthquakes occurred between 1985 and 1999 with magnitude less than 5.5.

Figure 3. Cumulative number of earthquakes vs. magnitude on a logarithmic scale. About 12000 earthquakes of low magnitude (less than 2.5) were used in this study. Data is represented by squares. The line is the cumulative number calculated from 6.